\documentclass[12pt]{amsart}
\usepackage{a4wide}\usepackage[pagebackref]{hyperref}
\usepackage{latexsym}\usepackage{amsbsy}\usepackage{amssymb}
\usepackage{amsmath}\usepackage{amsthm}\usepackage{amsfonts}
\usepackage{amstext}\usepackage{amsgen}\usepackage{amsbsy}
\usepackage{amsopn}\usepackage{amscd}
\usepackage{epsfig}\usepackage{graphicx}
\usepackage{mathrsfs,color}

\usepackage[latin1]{inputenc} 

\newtheorem{theorem}{Theorem}[section]
\newtheorem{lemma}[theorem]{Lemma}
\newtheorem{proposition}[theorem]{Proposition}

\newtheorem{definition}[theorem]{Definition}
\newtheorem{corollary}[theorem]{Corollary}

\newcommand{\dist}{\mathrm{dist}}

\newcommand{\eps}{\varepsilon}
\newcommand{\al}{\alpha}
\newcommand{\Om}{\Omega}
\newcommand{\Ga}{\Gamma}

\newcommand{\de}{\delta}
\newcommand{\cof}{\mathrm{cof}}\newcommand{\vphi}{\varphi}

\newcommand{\comp}{\subset\subset}\newcommand{\pa}{\partial}
\newcommand{\supp}{\mathrm{supp}}

\newcommand{\R}{\mathbb R}\newcommand{\N}{\mathbb N}
\newcommand{\Z}{\mathbb Z}
\newcommand{\A}{{\mathcal A}}

\newcommand{\Ll}{\ensuremath{L_{\mathrm{\scriptstyle loc}}}}
\newcommand{\Hl}{\ensuremath{H_{\mathrm{\scriptstyle loc}}}}
\newcommand{\Wl}{\ensuremath{W_{\mathrm{\scriptstyle loc}}}}
\newcommand{\D}{\ensuremath{{\mathcal D}}}
\newcommand{\G}{\ensuremath{{\mathcal G}}}
\newcommand{\gs}{\ensuremath{{\mathcal G}}}

\newcommand{\esm}{\ensuremath{{\mathcal E}_M}}
\newcommand{\ns}{\ensuremath{{\mathcal N}} }

\newcommand{\Cinfty}{\ensuremath{\mathcal C}^\infty}
\newcommand{\GT}{\Hl^1\cap\Ll^\infty}
\newcommand{\itemcr}{\hrule height 0pt width 40pt\hfill\break\vspace*{-\baselineskip}}
\newcommand{\g}{\ensuremath{\mathbf{g}}}
\newcommand{\GTconv}{\ensuremath{\Hl^1\cap\Ll^p}}

\newcommand{\Riem}{\mathrm{Riem}}
\newcommand{\Ric}{\mathrm{Ric}}
\newcommand{\DO}{\ensuremath{\mathcal P}}

\title{On the Geroch-Traschen class of metrics}
\author[R. Steinbauer]{Roland Steinbauer}
\address{Fakult\"at f\"ur Mathematik, Universit\"at Wien,
Nordbergstrasse 15, A-1090 Wien, Austria}
\email{Roland.Steinbauer@univie.ac.at}
\urladdr{http://www.mat.univie.ac.at/\~{}stein/}

\author[J. A. Vickers]{James A.\ Vickers}
\address{School of Mathematics, University of Southampton,
Highfield, Southampton SO17 1BJ, United Kingdom}
\email{J.A.Vickers@maths.soton.ac.uk}
\urladdr{http://www.soton.ac.uk/maths/}
\thanks{This work was supported by projects P16742, P20525,
 and Y-237 of the Austrian Science Fund.}
\date{October, 20 2008}

\begin{document}

\begin{abstract}
We compare two approaches to Semi-Riemannian metrics of low
regularity. The maximally ``reasonable'' distributional setting
of Geroch and Traschen is shown to be consistently contained in
the more general setting of nonlinear distributional geometry in the
sense of Colombeau.
\end{abstract}
\maketitle

\section{Introduction}

In this paper we deal with different approaches to metrics of low
differentiability in general relativity. While normally relativity is
formulated for smooth metrics, most of the relevant differential
geometric results actually hold in the case where the metric is only locally
$C^{2-}=C^{1,1}$, i.e., the first derivative being locally Lipschitz
continuous. In particular, this condition directly gives unique
(local) solvability of the geodesic equation. Moreover, by
Rademacher's theorem the second derivatives are in $\Ll^\infty$, hence
the Riemann tensor can be regarded as a distribution.

When further lowering the differentiability of the metric one meets
conceptual problems as one reaches the limits of classical (i.e.,
linear) distribution theory.  Since Einstein's equations are nonlinear,
one cannot simply pass from smooth solutions of the field equations to
weak ones. In particular, the curvature tensor is only linear in the
second derivatives of the metric but nonlinear in the lower order
terms. Hence one cannot simply calculate the curvature from a general
distributional metric.

In a classic paper \cite{gt} Geroch and Traschen studied the question
under which minimal conditions on the metric one can compute the
curvature. To be precise, they isolated a class of metrics---which we
will refer to as {\em gt-regular}---for which on the one hand one may
calculate the classical distributional curvature, and on the other
hand possesses a certain stability property. That is, they defined a
notion of convergence for gt-regular metrics which implies the
convergence of the respective curvature tensors in the class of
distributions. Note that it is this stability property which makes it sensible
to use gt-regular metrics to model singular matter configurations in
relativity.  A slightly more general class of metrics allowing for a
distributional curvature tensor but lacking stability in the above sense was
introduced by Garfinkle \cite{garfinkle}. Finally, we also mention
that the class of gt-regular metrics recently was rederived in a
coordinate-free manner in \cite{LFM}, see also \cite{srpski-proc}.

Although belonging to the Geroch-Traschen class is a sufficient
condition to allow one to compute the distributional curvature, the
question of necessity is more subtle. There are, however, indications
that the gt-regular metrics form the largest ``reasonable'' class of
distributional metrics: for example the only slightly more general
Garfinkle class fails to be stable, while even for gt-regular metrics
one cannot formulate the Bianchi identifies for example.

However Geroch and Traschen also proved that a gt-regular metric
allows only for a limited range of concentration of the gravitating
source: the curvature tensor of a gt-regular metric is supported on a
manifold of codimension of at most one. This explicitly excludes many
interesting scenarios, in particular, strings of matter and point
particles.

In order to model a wider class of spacetimes some authors were lead
to use alternative mathematical tools to describe space-times of low
regularity. In particular, the theory of algebras of generalised
functions due to J.F.\ Colombeau \cite{c1,c2,Cbull} proved to be
useful in the context of cosmic strings \cite{clarke,genhyp},
Kerr-Schild geometries \cite{hb5}, and impulsive pp-waves
\cite{herbertgeo,geo2}. Also it was used to study the initial value
problem for the wave equation in conical space-times \cite{genhyp} and
in singular space-times with locally bounded metrics \cite{waveq}; for
a recent overview see \cite{SV}. This approach goes beyond the limits
of classical (linear) distribution theory---hence also beyond the
class of gt-regular metrics---as it allows one to assign a product to an
arbitrary pair of distributions.  It is based upon regularising
distributions via convolution and the use of asymptotic estimates in
terms of a regularisation parameter. In many cases it also allows one to
compare the result of a calculation in the algebra of generalised
functions with classical distributions; this concept, called
association, basically consists in looking at the weak limit as the
regularisation parameter goes to zero.

In the case where we are given a gt-regular metric we therefore have two
approaches at hand to compute the curvature: the classical
distributional one due to Geroch and Traschen and the generalised
function approach using Colombeau's construction. The natural question
therefore arises as to whether these two approaches lead to the same
answer. In this paper we give a complete and positive answer to this
question.  Along the way we prove several results on convergence of
sequences of metrics generated via smoothing by convolution of
gt-regular metrics which are of interest in their own right and
provide refined stability results on the Geroch-Traschen class of metrics.

\section{Prerequisites}

In this section we introduce some notation and recall known material
on linear and nonlinear distributional geometry to make the
presentation self-contained. In particular, we define the notions of
gt-regular as well as generalised metrics and collect some basic
results on smoothings via convolution with strict delta nets.

We begin with some notational conventions. Throughout this paper
$\Omega$ denotes an open subset of $\R^n$ and $M$ an oriented, smooth
manifold of dimension $n$.  Given two subsets $U$ and $V$ of $\Om$ or
of $M$ we use the notation $V\comp U$ if the closure $\bar V$ of $V$
is still a subset of the interior $U^\circ$ of $U$.  Moreover, $K$ and
$L$ will always denote compact sets and $C$ will denote a generic
constant.

\subsection{Linear distributional geometry}
The space of distributions on $M$ is the dual space (in the sense of
the usual (LF)-topology) of the space of compactly supported
$n$-forms, i.e., $\D'(M)=(\Om^n_c(M))'$. Distributional sections of a
vector bundle $E\to M$ over $M$ are defined as elements of the dual space of the
compactly supported sections of $E^*\otimes \Lambda^n(M)$, where $E^*$
denotes the bundle dual to $E$ and $\Lambda^n(M)=T^*M\wedge\dots\wedge
T^*M$. Likewise distributional sections can be viewed as
$\Cinfty$-linear maps from the sections of the dual bundle $\Ga(E^*)$
to $\D'(M)$ or as sections of $E$ with distributional, coefficients,
that is we have
\begin{equation}\label{distgeo}
 \D'(E):=\big(\Ga(E^*\otimes\Lambda^n(M)\big)'
 \cong L_{\Cinfty(M)}\big(\Ga(E^*),\D'(M)\big)
 \cong \D'(M)\otimes_{\Cinfty(M)}\Ga(E).
\end{equation}
The space of distributional tensor fields (tensor distributions) of type 
$(r,s)$ is denoted ${\D'}^r_s(M)$. There is a well-developed theory which parallels
the smooth one but suffers from the natural limitations of
distribution theory, e.g.\ in all multilinear operations only one
factor may be distributional, while all others have to be
smooth~\cite{marsden,parker}.  For a pedagogical account
see~\cite[Sec.\ 3.1]{GKOS}.

Next we recall the definition of the (local) Sobolev spaces of integer
order, i.e., for $m\in\N_0$ and $1\leq p\leq\infty$ we set
\[
 W^{m,p}(\R^n):=\{u\in\D'(\R^n):\ \pa^\al u\in L^p(\R^n)\ \mbox{for all multi-indices with}\
 |\al|\leq m\}
\]
and denote the respective norms by $\|\ \|_{m,p}$. For any $\Omega$ we set
\[
 \Wl^{m,p}(\Om):=\{u\in\D'(\Om):\ \chi u\in W^{m,p}(\R^n)\ \mbox{for all test-functions}\
 \chi\in\D(\Om) \}.
\]
Note that $u\in\D'(\Om)$ is in $\Wl^{m,p}(\Om)$ iff on any open
$V\comp\Om$ it agrees with a function in
$W^{m,p}(\R^n)$. The space $\Wl^{m,p}(\Om)$ is a Fr\'echet space with its
topology induced by the family of semi-norms $p_\chi(u):= \|\chi
u\|_{m,p}$ or alternatively by the $\|\ \|_{m,p}$-norms on all
relatively compact subsets $V$ of $\Om$, which we denote by $\|\
\|_{W^{m,p}(V)}$.

On $M$ we define the local Sobolev spaces by means of local charts:
denote by $(U_\al,\phi^\al)$ the charts of some atlas for $M$, then we set
\[
 \Wl^{m,p}(M):=\{u\in\D'(M):\ \phi^{\al}_*u\in\Wl^{m,p}(\phi^\al(U_\al))\ \mbox{for all $\al$}\},
\]
where $\phi^\al_*$ denotes the push forward under the chart. $\Wl^{m,p}(M)$
is again a Fr\'echet space with its topology defined via the
semi-norms of $\phi^\al_*u$ in $\Wl^{m,p}(\phi^\al(U_\al))$, and one
may show that this definition does not depend on the atlas.  Finally,
for $E\to M$ one defines the space of $\Wl^{m,p}$-sections likewise
via vector bundle charts but for our purpose it will be sufficient to
think of them as sections with $\Wl^{m,p}$-coefficients, i.e.,
\[
\Wl^{m,p}(E) = \Wl^{m,p}(M)\otimes_{\Cinfty(M)}\Ga(E).
\]
In case $p=2$ we use the usual convention and write $\Hl^m$ for
$\Wl^{m,2}$ and in case $m=0$ we obtain the usual (local) Lebesgue
spaces which we denote by $\Ll^p$.

In \cite{gt} Geroch and Traschen defined the following class of metrics which we will
call gt-regular.
\begin{definition}[gt-regular metrics]\itemcr
\begin{enumerate}
\item We call a section of any vector bundle of regularity $\GT$ gt-regular.
\item A gt-regular metric $\g$ is a gt-regular section of  $T^0_2(M)$
which is a Semi-Riemannian metric (of fixed index) almost everywhere.
\end{enumerate}
\end{definition}
The motivation for Geroch and Traschen to introduce this notion is
that it follows from the coordinate definition that for a gt-regular
metric it is possible to give a distributional definition of the
Riemannian curvature tensor.

\subsection{Smoothings}
Next we recall the convergence properties of smoothing via
convolution. The mollifiers we are going to use will be slightly more
general than the standard ones obtained by scaling one fixed
test-function with unit integral.  More precisely we shall use.
\begin{definition}[Smoothing with strict delta nets]\itemcr
\begin{itemize}
\item [(1)] A net $(\psi_\eps)_{\eps\in(0,1]}$ of smooth functions on $\R^n$ 
is called a strict delta net, if
\begin{enumerate}
\item $\supp(\psi_\eps)\to\{0\}$ for $\eps\to 0$
\item $\int\psi_\eps\to 1$ for $\eps\to 0$
\item $\psi_\eps$ is uniformly bounded in $L^1$, i.e., $\exists\ C_\psi:\ \|\psi_\eps\|_{L^1}
 \leq C_\psi$ for all $\eps$.
\end{enumerate}
\item [(2)] For any strict delta net $(\psi_\eps)_\eps$ we denote by
$d_{\psi_\eps}$ the diameter of the support of $\psi_\eps$, i.e.,
$d_{\psi_\eps}:=sup\{|x|:\ x\in\supp(\psi_\eps)\}.$
\item [(3)] For any $f\in\Ll^1(\Om)$ we call the convolution $f_\eps$ of $f$ with a
strict a delta net $(\psi_\eps)_\eps$ a \emph{smoothing} of $f$, i.e., for
$x\in\Om_{\psi_\eps}:=\{y\in\Om:\ \dist(y,\pa\Om)> d_{\psi_\eps}\}$ we set
\[f_\eps(x):=f*\psi_\eps(x)=\int\limits_{B(x,d_{\psi_\eps})}f(x-y)\psi_\eps(y)\,dy,\]
where $B(x,r)$ denotes the open ball of radius $r$ around $x$.
\end{itemize}
\end{definition}

We recall the following results on smoothings (which are a mild
generalisation of the ones found e.g.\ in \cite[\S5.3, \S
C.4]{evans}).
\begin{lemma}[Smoothing via convolution]\label{lemma:smoothing}
The smoothing of any $f\in \Ll^1(\Om)$ has the following properties.
\begin{enumerate}
\item $f_\eps\in\Cinfty(\Om_{\psi_\eps})$ and $f_\eps\to f$ almost everywhere.
\item If $f$ is continuous the convergence is actually uniform on compact subsets of $\Om$.
\item\label{Wloc} If $f\in\Wl^{m,p}(\Om)$ for $1\leq p<\infty$ then $f_\eps\to f$ in $\Wl^{m,p}(\Om)$
\end{enumerate}
\end{lemma}

Note that for $f\in\Ll^\infty(\Om)$ the last item implies $f_\eps\to
f$ in $\Ll^p(\Om)$ for all $p<\infty$ but not $p=\infty$. Indeed, the
latter would contradict non-separability of $L^\infty$.  For later
reference we remark that also in this case $f_\eps$ is nevertheless
locally uniformly bounded. More precisely, we have for all $1\leq
p\leq\infty$ and all $f\in\Ll^p(\Om)$ that for any $V\comp\Om$
\begin{equation}\label{boundedness}
\|f_\eps\|_{L^p(V)}\leq \|\psi_\eps\|_{L^1}\|f\|_{L^p(W)}\leq C_\psi\,\|f\|_{L^p(W)},
\end{equation}
where $W$ is any relatively compact subset of $\Om$ with $V\comp W$.

\subsection{Nonlinear distributional geometry}\label{colombo}
In nonlinear distributional geometry \cite{ndg}, \cite[Ch.\ 3]{GKOS} (in the sense
of J.F.\ Colombeau\cite{c1,c2,Cbull}) one replaces the vector space $\D'(M)$
of distributions by the algebra of generalised functions $\G(M)$ to
overcome the problem of multiplication of distributions.

The basic idea of the construction is smoothing of distributions (via
convolution) and the use of asymptotic estimates in terms of a
regularisation parameter. The (special) Colombeau algebra of
generalised functions on $M$ is defined as the quotient
$$\gs(M) := \esm(M)/\ns(M)$$
of moderate nets of smooth functions modulo negligible ones, where
the respective notions are defined by
\[
\esm(M) :=\{ (u_\eps)_\eps\in\Cinfty(M):\, \forall K\ \mbox{compact}\,
\forall P\in\DO(M)\, \exists N \in \N:\, \sup\limits_{p\in K}|Pu_\eps(p)|=O(\eps^{-N}) \}\\
\]
\[\ns(M)  :=\{ (u_\eps)_\eps\in\Cinfty(M):\, \forall K\ \mbox{compact}\,
\forall P\in\DO(M)\, \forall m \in \N:\, \sup\limits_{p\in K}|Pu_\eps(p)|=O(\eps^{m}) \},
\]
with $\DO(M)$ denoting the space of linear differential operators on
$M$.  Elements of $\gs(M)$ are denoted by $u = [(u_\eps)_\eps] =
(u_\eps)_\eps + \ns(M)$. With componentwise operations, $\gs(M)$ is a
fine sheaf of differential algebras where the derivations are Lie
derivatives with respect to classical vector fields defined according
to the formula $L_Xu :=[(L_X u_\eps)_\eps]$. The spaces of moderate
resp.\ negligible sequences and hence the algebra itself may be
characterised locally, i.e., $u \in \gs(M)$ iff $\phi^\al_*u \in
\gs(\phi^\al(U_\al))$ for all charts $(U_\al, \phi^\al)$, where, on
the open set $\phi^\al(U_\al) \subset \R^n$, partial derivatives
replace differential operators in the respective estimates.

The $\gs(M)$-module of generalised sections in $E\to M$ can be defined
along the same lines using analogous asymptotic estimates.  However,
as in the case of $\Wl^{mp}$-valued sections it is more convenient to
use the following algebraic description of generalised tensor fields
\[ \gs(E) = \gs(M) \otimes \Ga(E).\]
Hence generalised tensor fields are just given by classical ones with
generalised coefficient functions. Moreover, we have the following
chain of isomorphisms
\begin{equation}\label{gengeo}
\gs(E) \cong L_{\Cinfty(M)}(\Ga(E^*),\gs(M))\cong L_{\gs(M)}(\G(E^*),\gs(M)).
\end{equation}
Spaces of generalised tensor fields will be denoted by $\G^r_s(M)$.
Note that in contrast to classical distributions (c.f.\ (\ref{distgeo})),
generalised sections map generalised (and not merely smooth) sections
of the dual bundle to generalised functions. It is precisely this
property that allows one to raise and lower indices with the help of a
generalised metric (see below) just as in the smooth setting.

Smooth functions are embedded into $\gs(M)$
simply by the \lq\lq constant\rq\rq\ embedding $\sigma$, i.e.,
$\sigma(f) := [(f)_\eps]$. On $\Om$ compactly
supported distributions are embedded into $\gs$ via convolution with
a mollifier $\rho \in \mathcal{S}(\R^n)$ with unit integral
satisfying $\int \rho(x) x^\alpha dx = 0$ for all $|\alpha| \geq 1$;
more precisely setting $\rho_\eps(x) = (1/\eps^n) \rho(x/\eps)$, we
have $\iota_0(w) := [(w * \rho_\eps)_\eps]$. (The fact that all moments of
$\rho$ vanish is used to prove that $\iota_0|_{\Cinfty}=\sigma$
in the quotient, which implies that the product of smooth functions is preserved
in the construction---a distinguished feature of this approach, see below.)
In case $\supp(w)$ is non-compact (hence $w*\rho_\eps$ is not defined),
one employs a sheaf-theoretic construction (\cite[Sec.\ 1.1.2]{GKOS}) or
alternatively uses an additional cut off at a different rate of growth (cf.\
\cite{del}): We set $\psi_\eps(x):=\chi(x/\sqrt{\eps})\rho_\eps(x)$ where
$\chi\in\D(B_2(0))$, $\chi=1$ on $B_1(0)$ and
\[
 \D'(\Om)\ni w\mapsto \iota_\psi(w):=[(w*\psi_\eps)_\eps]\in\G(\Om)\
 \mbox{($\eps$ small enough)},
\]
to obtain an embedding of distributions by convolution with
a strict delta net which obviously commutes with derivatives.
Note that this construction depends on the choice of the
mollifier $\rho$ (but not on $\chi$), which allows for a flexible way of modelling
singular objects. Moreover this construction can be lifted to $M$
decomposing $w$ via a partition of unity subordinate to the charts of an atlas and
chartwise convolution (cf.\ \cite[Thm.\ 3.2.10]{GKOS}). Such a procedure is,
of course, dependent of the choice of charts and partition functions, hence non-geometric
in an essential sense. There is, however, a version of the construction possessing a
canonical and invariant embedding of $\D'(M)$ resp.\ ${\D'}^r_s(M)$, the so-called full
Colombeau algebras, see~\cite{GlobTh}, resp.~\cite{GlobTh2}. For the purpose of the present work
it is, however, more convenient to use the (technically less demanding) special version: In fact
we are going to derive convergence results for embedded distributions in the $\Wl^{m,p}$- and
$\D'$-topologies respectively, which take place on (relatively) compact sets and so
will be independent of the choice of charts, partition functions and also of $\rho$.

Finally, in light of Schwartz' impossibility result~\cite{Schw1},
the setting introduced above gives a minimal framework within which
sections of vector bundles, and, in particular, tensor fields may be
subjected to nonlinear operations, while maintaining consistency with
smooth and distributional geometry: tensor products of smooth sections are
preserved as well as derivatives of distributional sections.

The interplay between generalised functions and distributions is most
conveniently formalised in terms of the notion of
association. A generalised function $u \in \gs(M)$ is called
associated to zero, $u \approx 0$, if one (hence any) representative
$(u_\eps)_\eps$ converges to zero weakly. The equivalence relation $u
\approx v :\Leftrightarrow u - v \approx 0$ gives rise to a linear
quotient of $\gs$ that extends distributional equality. Moreover, we
call a distribution $w \in \D'(M)$ the distributional shadow
or macroscopic aspect of $u$ and write $u \approx w$ if, for
all compactly supported $n$-forms $\boldsymbol{\nu}$ and one (hence
any) representative $(u_\eps)_\eps$, we have
\[
\lim_{\eps \to 0} \int\limits_M u_\eps \boldsymbol{\nu} = w(\boldsymbol{\nu}).
\]
By~\eqref{gengeo}, embeddings and association extend to
generalised sections in a natural way.

Finally we recall the basic notions of Semi-Riemannian geometry in the
generalised setting.
\begin{definition}[Generalised metric]\label{gen-metric}
A symmetric section $\g\in\G^0_2(M)$ is called a generalised
Semi-Riemannian metric if $\det\g$ is invertible in $\G(\mathrm{Vol}^2(M))$, i.e.,
for any representative
$\left( \det( \mathbf{g}_{\eps}) \right)_\eps$ of $\det\mathbf{g}$ we have
\[
\forall K\ \mbox{compact}\ \exists m \in \N:\
\inf_{p\in K} |\det( \mathbf{g}_{\eps})| \geq \eps^m.
\]
\end{definition}
Here $\mathrm{Vol}^q(M)$ denotes the bundle of $q$-densities on $M$.
The following characterisation of generalised metrics captures the
intuitive idea of a generalised metric as a net of classical metrics
approaching a singular limit: $\mathbf{g}$ is a generalised metric iff
on every relatively compact open subset $V$ of $M$ there exists a
representative $( \mathbf{g}_{\eps})_\eps$ of $\mathbf{g}$ such that,
for fixed $\eps$, $\mathbf{g}_{\eps}$ is a classical metric and its
determinant, $\det\g$, is invertible in the generalised sense. The
latter condition basically means that the determinant is not too
singular.

A generalised metric induces a $\gs(M)$-linear isomorphism from
$\gs^1_0(M)$ to $\gs^0_1(M)$. The inverse of this isomorphism gives
a well-defined element of $\gs^2_0(M)$, the inverse metric, which we
denote by $\mathbf{g}^{-1}$, with representative
$\left( \mathbf{g}_{\eps}^{-1} \right)_\eps$. The
generalised covariant derivative, as well as the generalised
Riemann-, Ricci- and Einstein tensors, of a generalised metric is
defined by the usual formulae at the level of representatives. For
further details see~\cite{KS1} and \cite[Sec.\ 3.2]{GKOS}.

\section{Notions of nondegeneracy}
In this short section we discuss some notions of nondegeneracy for metrics of
low differentiability.

In the purely distributional setting, that is considering a metric as
a symmetric element of $\D'^0_2(M)$, one finds two different notions
of nondegeneracy in the literature, i.e.,
\begin{itemize}
\item [(A)] Marsden in \cite{marsden} defines $\g$ to be nondegenerate if $\g(X,Y)=0$
for all smooth vector fields $X$, implies that the smooth vector field $Y$ vanishes.
\item [(B)] Parker in \cite{parker} defines $\g$ to be nondegenerate if it is nondegenerate
off its singular support.
\end{itemize}
Note that notion (A) is strictly weaker than the usual pointwise condition. For example
\begin{equation}\label{degmet}
 ds^2=x^2dx^2
\end{equation}
on $\R$ is nondegenerate in the sense of Marsden but is clearly not
invertible on the whole of $\R$. On the other hand condition (B) does
not put any restrictions on $g$ at the points where $g$ is not
smooth. So the best option for a distributional metric would be to
call it nondegenerate if both (A) {\em and} (B) hold.

On the other hand, in our view, the notion of nondegeneracy for
gt-regular metrics was not unambiguously defined in
\cite{gt}. The original statement saying that ``the inverse of the
metric exists everywhere'' is mathematically best interpreted by
saying that in the $\Ll^\infty\cap\Hl^1$-class of $\g$ there exists a
representative which is invertible everywhere. This, however, would
allow metric (\ref{degmet}) to again count as nondegenerate: simply set the
coefficient equal to $1$, for example, at $x=0$.

However, a natural notion of nondegeneracy for gt-regular metrics is
available (see also~\cite{srpski-proc}). Note that the space $\GT$ is
actually an algebra. Indeed, $fg$ clearly is in $\Ll^\infty\subseteq\Ll^2$ 
and to show that $\pa_j(fg)\in\Ll^2$ we use the Leibnitz rule (which applies
in all $\Wl^{1,p}$, $p\geq2$)
to write $\pa_j(fg)=(\pa_j f)g+(\pa_jg)f$ which is a sum of products 
$\Ll^2\times\Ll^\infty$ hence in $\Ll^2$. Also a function $f\in\GT$ which
is {\em locally uniformly bounded away from zero}, i.e., which satisfies
\begin{equation}\label{lubafz}
 \forall K\ \mbox{compact}\ \exists C_K:\ |f(x)|\geq C_K>0\
 \mbox{almost everywhere on}\ K,
\end{equation}
is invertible and $1/f\in\GT$ is again locally uniformly bounded away
from zero.  Therefore we employ the following definition of
nondegeneracy for gt-regular metrics (see also \cite{LFM}, p.\ 14).
\begin{definition}[Nondegeneracy of gt-regular metrics]\label{gt-nondeg}\quad \\
We call a gt-regular metric $\g$ nondegenerate if its determinant is locally
uniformly bounded away from zero, i.e.,
\begin{equation}\label{nondeg}
 \forall K\ \mbox{compact}\ \exists C_K:\ |\det\g(x)|\geq C_K>0\
 \mbox{almost everywhere on}\ K.
\end{equation}
\end{definition}
Hence the determinant $\det\g$ of a nondegenerate gt-regular metric
$\g$ is an invertible density of regularity $\GT$ with
$(\det\g)^{-1}\in\GT$ again locally uniformly bounded away from
zero. Hence by the cofactor formula the inverse $\g^{-1}$ of $\g$ is
again of regularity $\GT$ and nondegenerate in the sense that its
determinant $\det(\g^{-1})$ is locally uniformly bounded away from
zero.

However, this notion of nondegeneracy still does not have optimal
stability properties with respect to smoothing via convolution and we
will come back to discuss this issue in section~\ref{smoothing gt}.

To end this section we remark that the problems discussed above all
originate from the fact that neither the distributional nor the
gt-setting can provide pointwise resp.\ pointwise everywhere control
on the metric. In contrast to this the condition of nondegeneracy
employed for generalised metrics in Definition~\ref{gen-metric} allows
for a pointwise control on generalised points as is shown
in~\cite[Thm.\ 3.2.4]{GKOS}.

\section{Smoothing gt-regular metrics}\label{smoothing gt}

In this section we provide a detailed account on stability properties of
gt-regular metrics under smoothing with strict delta nets and
of convergence results of embeddings of gt-regular metrics into the Colombeau
algebra.

We introduce the following notation: given a gt-regular metric $\g$
with local components $g_{ij}$ we will write $g^\eps_{ij}$ for their
smoothings, i.e., $g^\eps_{ij}=g_{ij}*\psi_\eps$, with $(\psi_\eps)_\eps$ being 
a strict delta net, and denote the resulting metric by $\g_\eps$.  

To begin with we collect together some convergence results for products of nets
of functions in $\GT$ generated by smoothing via convolution with
strict delta nets.
Given a function $f\in\GT$ we have from Lemma~\ref{lemma:smoothing}
(iii) that $f_\eps\to f\in\Hl^1\cap\Ll^p$ for all $p<\infty$. Also
given $f_1,\dots,f_m\in\GT$ the product $f_1\cdots f_m$ is in $\GT$
and $(f_1\cdots f_m)_\eps=(f_1\cdots f_m)*\psi_\eps\to f_1\cdots f_m$
in $\Hl^1\cap\Ll^p$ for all $p<\infty$.  We shall, however, be
interested in convergence of curvature quantities derived from the
componentwise smoothing of gt-regular metrics. Hence we have to
study convergence properties of (derivatives of) $f_{1\,\eps}\cdots
f_{m\,\eps}$ rather than $(f_1\cdots f_m)_\eps$.

Next we connect the products of nets of functions that arise in our
approach to the general theory given in \cite[Sec.\ II.7]{MOBook}.  In
the latter context the product we are dealing with is called strict
product (of type (7.4)), that is, given two distributions $u$ and $v$
we look at the limit
\begin{equation}
\lim_{\eps\to 0}(u*\psi_\eps)(v*\psi_\eps).
\end{equation}
If it exists for all strict delta nets $(\psi_\eps)_\eps$ (it is then
automatically independent of the particular choice of $\psi_\eps$) we
call the limit the strict product of $u$ and $v$ and denote it by
$[uv]$. The strict product can be placed in a hierarchy of products of
distributions (see \cite{MOBook}, p.\ 69) which are all compatible
with the Colombeau product in the sense of association (\cite{MOBook},
Prop.\ 10.3).

Likewise we can make use of the $\Wl^{m,p}$-duality product, that is
also contained in the above mentioned hierarchy. More precisely, one
can define (by duality, \cite{MOBook}, Prop.\ 5.2) a continuous
product
\begin{equation}
\Wl^{m,q}\times\Wl^{l,p}\to\Wl^{k,r}
\label{sobolev}
\end{equation}
if $l,m\in\Z$ with $l+m\geq 0$, $1\leq p,q\leq\infty$ with
$1/p+1/q\leq1$ and $k,r$ are defined by $k:=\min(l,m)$,
$1/r:=1/p+1/q$.  (For the spaces $\Wl^{m,p}$ for negative $k$ see
e.g.\ \cite[Ch.\ 3]{adams}---although we will not need to consider
them in the following.) This product is more special than the strict
product, although it is compatible with it, and has the additional
benefit that it is partially associative, i.e., $(fu)v=u(fv)$ for all
smooth $f$.  We will have to deal with products containing many
factors and so we have to be careful with the loss in $r$ in formula
\eqref{sobolev}; only the special case $p=q=\infty$ does not share
this problem but, on the other hand this case, lacks stability under smoothing
as discussed below Lemma~\ref{lemma:smoothing}.

We now give a useful auxiliary result which (partially) follows from the general
statements above and is needed to establish the results later in this
section.
\begin{lemma}[Convergence of products in $\GT$]\label{basic-conv}
Let $(\psi_\eps)_\eps$ be a strict delta net.
\begin{enumerate}
\item If $g_1,\dots g_m\in\Ll^\infty$ $(m\in\N)$, then
\[
 g_{1\,\eps}\cdots g_{m\,\eps}\to g_1\cdots g_m\ \mbox{in}\ \Ll^p\ \mbox{for all}\ p<\infty.
\]
\item If $f\in\Ll^p$ with $1\leq p<\infty$ and $(g_\eps)_\eps$ is a locally uniformly bounded net
converging pointwise almost everywhere to some $g\in\Ll^\infty$, then
\[
 f_\eps g_\eps\to fg\ \mbox{in}\ \Ll^p.
\]
\item If $f_1,\dots,f_m\in\GT$ $(m\in\N)$, then
\[
 f_{1\,\eps}\cdots f_{m\,\eps}\to f_1\cdots f_m\ \mbox{in}\ \GTconv\
 \mbox{for all $p<\infty$}.
\]
\end{enumerate}
\end{lemma}

Observe that statement (iii) says that the product of the smoothings
of gt-regular functions converges in the same sense (i.e., in
$\GTconv$ for all $p<\infty$) as the smoothing of (a product of)
gt-regular functions.

\begin{proof} (i) On any relatively compact set $V$ we have for all $p<\infty$
\begin{eqnarray*}
\|g_{1\,\eps}\cdots g_{m\,\eps}-g_1\cdots g_m\|_{L^p(V)}
&\leq&\dots+\|g_{1\,\eps}\cdots(g_{j\,\eps}-g_j)\cdots g_m\|_{L^p}+\dots\\
&\leq&\dots+\|g_{1\,\eps}\cdots g_{j-1\,\eps}g_{j+1}\cdots g_m\|_{L^\infty}
 \|g_{j\,\eps}-g_j\|_{L^p}+\dots\quad.
\end{eqnarray*}
Now the respective first terms are bounded by estimate (\ref{boundedness}) and the convergence
is due to Lemma~\ref{lemma:smoothing}(\ref{Wloc}).

\noindent (ii)
On any $V$ as above we write
\[
 \|f_\eps g_\eps-fg\|_{L^p(V)}\leq
 \|g_\eps\|_{L^\infty}\|f_\eps-f\|_{L^p}+\|f(g_\eps-g)\|_{L^p}.
\]
For the first term convergence follows from Lemma~\ref{lemma:smoothing}(\ref{Wloc})
and the assumptions on $(g_\eps)_\eps$. To deal with the other term observe that
\begin{equation}\label{domconv}
\mbox{$fg_\eps\to fg$ in $L^p(V)$}.
\end{equation}
Indeed we have convergence almost everywhere by assumption as well as
$|f(x)g_\eps(x)|\leq C|f(x)|\in L^p(V)$ almost everywhere.
So dominated convergence applies to give the result.

\noindent (iii)
$\Ll^p$-convergence for all $p<\infty$ follows from (i) and we only have to show
$\Ll^2$-convergence of the derivatives.
By the Leibnitz rule we have to show that for all $1\leq j\leq n$ and all $1\leq l\leq m$
\[f_{1\,\eps}\cdots f_{l-1\,\eps}(\pa_j f_{l\,\eps})f_{l+1\,\eps}\cdots f_{m\,\eps}
 \to f_1\cdots f_{l-1}(\pa_j f_l)f_{l+1}\cdots f_m
\]
in $\Ll^2$. This, however, follows from (ii) with $p=2$, since $f_{1\,\eps}\cdots f_{l-1\,\eps}f_{l+1\,\eps}\cdots f_{m\,\eps}$ is a locally uniformly bounded net
which by Lemma~\ref{lemma:smoothing}(i) converges pointwise almost everywhere
to $f_1\cdots f_{l-1}f_{l+1}\cdots f_m$ and $(\pa_jf_{l\,\eps})=(\pa_jf_l)_\eps$
is the smoothing of an $\Ll^2$-function.
\end{proof}

We now obtain as a direct consequence of Lemma~\ref{basic-conv}(iii) a
stability result for the determinant of gt-regular metrics.
\begin{proposition}[Stability of the determinant]\label{det-conv}
Let $\g$ be a nondegenerate gt-regular metric and let $(\psi_\eps)_\eps$ be a strict delta net.
Then we have for the determinant of the smoothing
\[\det(\g_\eps)\to\det\g\ \mbox{in}\ \GTconv\ \mbox{for all}\ p<\infty.\]
In particular, we have for any embedding $\det(\iota (g))\approx\det (g)$.
\end{proposition}

Next we discuss nondegeneracy of the smoothing of a nondegenerate
gt-regular metric.  Of course, the key is that the determinant of the
smoothed metric has to be nonvanishing in an appropriate sense, which
turns out to be a delicate matter: Unfortunately
Proposition~\ref{det-conv} does not give pointwise (let alone uniform)
control on the determinant of the smoothing. Recall that such a
condition will be needed to prove that the smoothing of a
nondegenerate gt-regular metric is a generalised metric (cf.\
Definition~\ref{gen-metric}---we will aim at proving this condition
for $m=0$).

As a preparation we first discuss the scalar case. Suppose $f\in\GT$
is positive a.e.\ and locally uniformly bounded away from zero, i.e.,
satisfies (\ref{lubafz}). Then we know that $1/f\in\GT$ and we want to
secure that $1/f_\eps \to 1/f$ in $\GTconv$ for all $p<\infty$.
This will be achieved if $1/f_\eps$ is a uniformly bounded net on all
relatively compact $V$ for small $\eps$, which in turn is guaranteed
by the following condition
\begin{equation}\label{nze}
\forall K\ \mbox{compact}\ \exists C'_K\ \exists\eps_0(K):
f_\eps(x)\geq C'_K>0\ \forall x\in K,\ \forall \eps\leq\eps_0(K).
\end{equation}
which gives uniform control on the positivity of the smoothing.
Unfortunately this condition does not follow from (\ref{lubafz}) if we
use arbitrary strict delta nets. (As an explicit counterexample take
$f(x)=H(-x)+3H(x)$ with $H$ denoting Heaviside's step function and use
the strict delta net
$\psi_\eps(x)=1/\eps(2\rho((x-\eps)/\eps)-\rho((x+\eps)/\eps))$, where
$\rho$ is a standard bump function around zero with unit
integral. Then $f_\eps(0)=-1$ for all $\eps$.)

Indeed to preserve positivity in the above sense during the
smoothing (i.e., such that positivity and (\ref{lubafz}) imply (\ref{nze})) one would, in a first
attempt, use {\em positive} strict delta nets. However, recall that in the Colombeau
approach it is essential to use mollifiers with vanishing moments and that such a mollifier
cannot be positive. Nevertheless it is possible to provide strict delta nets which have
vanishing moments and at the same time allow for  good control on the $L^1$-norm of their
negative parts, which is the essential ingredient for preserving positivity. More
precisely we have the following result which we prove in the appendix.
\begin{lemma}[Existence of admissable mollifiers]\label{exmol}
 There exist strict delta nets $(\rho_\eps)_\eps$ with
 \begin{enumerate}
 \item  $\displaystyle\supp(\rho_\eps)\subseteq B_\eps(0)$\quad for all $\eps\in(0,1]$
 \item  $\displaystyle\int\psi_\eps(x)\, dx=1$\quad for all $\eps\in(0,1]$
 \end{enumerate}
 which are moderate, have finally vanishing moments and the negative
 parts have arbitrarily small $L^1$-norm, i.e., $(\rho_\eps)_\eps$ additionally satisfies
 \begin{itemize}
  \item [(iii)] $\displaystyle\forall\alpha\in\N_0^n\ \exists p:\
              \sup_{x\in\R^n}|\partial^\alpha\rho_\eps(x)|=O(\eps^{-p})$
  \item [(iv)] $\displaystyle\forall j\in\N\ \exists\eps _0:\
              \int x^\alpha\rho_\eps(x)\,dx=0$\quad for all $1\leq|\alpha|\leq j$ and all $\eps\leq\eps_0$
  \item [(v)] $\displaystyle\forall \eta>0\ \exists \eps_0:\
              \int|\rho_\eps(x)|\,dx\leq 1+\eta$\quad for all $\eps\leq\eps_0$.
 \end{itemize}
\end{lemma}

We will call strict delta nets $(\rho_\eps)_\eps$ as provided by
Lemma~\ref{exmol} {\em admissible mollifiers} and from now on consider
smoothings generated by convolution with such delta nets. Also,
convolution with an admissible strict delta net provides an embedding
$\iota_\rho$ of distributions in the Colombeau algebra as is shown in
Corollary~\ref{ca1} in the appendix.

We next show that smoothing with admissible mollifiers indeed
preserves positivity in an appropriate sense.
\begin{lemma}[Positive smoothing and convergence of the inverse]\label{pos-smooth}
Let $f\in\Ll^\infty$, $f>0$ almost everywhere and locally uniformly bounded away from zero, i.e.,
\[\forall K\ \mbox{compact}\ \exists C_k:\ f(x)\geq C_K>0\ \mbox{almost everywhere on $K$}.\]
Then for any admissible mollifier $(\rho_\eps)_\eps$ we have.
\begin{enumerate}
\item The smoothing $f_\eps=f*\rho_\eps(x)$ is a net, locally uniformly bounded away from zero, i.e.,
\[ 
\forall L\ \mbox{compact}\ \exists C'_L\ \exists \eps_0(L):\ f_\eps(x)
\geq C'_L>0\ \forall x\in L,\ \forall \eps\leq\eps_0(L).
\]
\item For any open and relatively compact set $V$ there exists $\eps_0(V)$ such
that $1/f_\eps$ is a smooth and uniformly bounded net on $V$, i.e., $\|1/f_\eps\|_{L^\infty(V)}\leq C$
 for all $\eps\leq\eps_0(V)$ and
\[
\frac{1}{f_\eps(x)}=\frac{1}{f*\rho_\eps(x)}\to \frac{1}{f}\
\mbox{in}\ \Ll^p\ \mbox{for all $p<\infty$}.
\]
\item If, in addition, $f\in\GT$ then we can strengthen the convergence assertion to
\[
\frac{1}{f_\eps(x)}=\frac{1}{f*\rho_\eps(x)}\to \frac{1}{f}\
\mbox{in}\ \GTconv\ \mbox{for all $p<\infty$}.
\]
\end{enumerate}
\end{lemma}

Note that although $1/f\in\GT$ even if we drop the positivity assumption in (ii) and only ask
for (\ref{lubafz}) the convergence result fails in general: such a function could change
sign forcing the smoothing to attain a zero.

\begin{proof}
(i) Let $L$ be compact, $x\in L$ and choose $K$ compact such that $L\comp K$. We split
$\rho_\eps$ into its positive and negative part (i.e., $\rho_\eps=\rho_\eps^+-\rho_\eps^-$,  with $\rho_\eps^+:=\max(\rho_\eps,0)$, $\rho_\eps^-:=-\min(\rho_\eps,0)$) to obtain
\begin{equation}\label{split}
f_\eps(x)=f*(\rho_\eps^+-\rho_\eps^-)(x)\geq f*\rho_\eps^+(x)-\|f*\rho_\eps^-(x)\|_{L^\infty(L)}.
\end{equation}
Estimating the first term on the r.h.s. of (\ref{split}) we have for $\eps$ small enough
\[
f*\rho_\eps^+(x)=\int f(x-y)\rho_\eps^+(y)\,dy \geq C_K\|\rho_\eps^+\|_{L^1}\geq C_K(1-\frac{\eta}{2}),
\]
where $\eta$ is the constant of Lemma~\ref{exmol}(v) and will be chosen appropriately later.
On the other hand we use inequality (\ref{boundedness}) on the second term on the
r.h.s. of (\ref{split}) to obtain
\[\|f*\rho_\eps^-\|_{L^\infty(L)}
\leq\|f\|_{L^\infty(K)}\|\rho_\eps^-\|_{L^1}\leq\|f\|_{L^\infty(K)}\frac{\eta}{2}.
\]
Combining the latter two estimates and choosing $\eta\leq C_K/(\|f\|_{L^\infty(K)}+C_K)$ we obtain
\[
f_\eps(x)\geq C_K(1-\frac{\eta}{2})-\|f\|_{L^\infty(K)}\frac{\eta}{2}\geq \frac{C_K}{2}=:C'_L>0.
\]

\noindent
(ii) Let $V$ be open and relatively compact. Then by (i) $1/f_\eps\in\Cinfty(V)$ form a
 uniformly bounded net for $\eps$ small enough. Moreover, $f_\eps\to f$ in $\Ll^p$
for all $p<\infty$ by Lemma~\ref{lemma:smoothing}(\ref{Wloc}).
So we find for all $p<\infty$
\[
\left\|\frac{1}{f_\eps}-\frac{1}{f}\right\|_{L^p(V)}
\leq\frac{\|f-f_\eps\|_{L^p}}{\|f_\eps f\|_{L^\infty}}
\leq\frac{1}{C_{\bar V}C'_{\bar V}}\,\|f_\eps-f\|_{L^p}
\to 0.
\]

\noindent
(iii)
In view of (ii) it remains to deal with the derivatives and we write for all $1\leq j\leq n$
\begin{eqnarray*}
\lefteqn{
\left\|\pa_j\Big(\frac{1}{f_\eps}-\frac{1}{f}\Big)\right\|_{L^2(V)}
=\left\|\frac{f^2\pa_jf_\eps-f_\eps^2\pa_jf}{f_\eps^2f^2}\right\|_{L^2}
\leq \frac{1}{ \|f^2_\eps f^2\|_{L^\infty} }\ \|f^2\pa_j f_\eps-f_\eps^2\pa_j f\|_{L^2}}\\
&\leq&\frac{1}{C_{\bar V}^2{C'}_{\bar V}^2}
\Big(
\|f^2\|_{L^\infty}\|\pa_j f_\eps-\pa_jf\|_{L^2}
 +\big(\|f\|_{L^\infty}+\|f_\eps\|_{L^\infty}\big)\|(f-f_\eps)\pa_j f\|_{L^2}\Big)
\end{eqnarray*}
Now the first term is converges to zero by Lemma~\ref{lemma:smoothing}(\ref{Wloc}) and
the second by dominated convergence (cf.\ (\ref{domconv}) for $p=2$).
\end{proof}

Now we return to the issue of nondegeneracy of the smoothings of
gt-regular metrics and take a close look at the determinant of the
smoothing.  Note that we have to deal with $\det(\g_\eps)$ rather than
$(\det\g)_\eps$, which means that we cannot simply use the results on
the scalar case above.
We again aim at some uniform control, more precisely at a condition of the form
\begin{equation}\label{det-pos}
\forall K\ \mbox{compact}\ \exists C'_K\ \exists \eps_0(K):\ |\det(\g_\eps)|\geq C'_K>0\
\forall x\in K,\ \forall \eps\leq\eps_0(K),
\end{equation}
since it will also imply that the smoothed metric is nondegenerate in
the generalised sense.  Of course if $\g$ was continuous then the
convergence would be locally uniform and the determinant $\det(\g_\eps)$
would obey (\ref{det-pos}) due to (\ref{nondeg}).  However, in the
general case we shall use the following {\em stability condition} for
gt-regular metrics.

\begin{definition}[Stability condition for gt-regular metrics]\label{gt-stab}
Let $\g$ be a gt-regular metric and denote by $\lambda_i,\dots,\lambda_n$ its eigenvalues.
\begin{enumerate}
\item For any compact $K$ we denote by
\[\mu_K:=\min\limits_{1\leq i\leq n}\mathop{\mathrm{ess\,inf}}\limits_{x\in K}|\lambda^i(x)|,\]
the (essential) absolute infimum of any eigenvalue of $\g$ on $K$.
\item We call $\g$ stable if for each compact $K$ there is a continuous $(0,2)$-tensor field
${\mathbf A}^K$ on $K$ such that for $1\leq j,k\leq n$
\begin{equation}\label{stab}
\mathop{\mathrm{ess\,sup}}\limits_{x\in K}|g_{ij}(x)-A_{ij}^K(x)|\leq C''_K<\frac{\mu_K}{2n}.
\end{equation}
\end{enumerate}
\end{definition}

Note that if $\g$ is nondegenerate then $\mu_K>0$ for all $K$. Then
the stability condition means that on compact sets the entries of the
metric $\g$ differ from those of a continuous function by an amount
proportional to the smallest eigenvalue, i.e, the entries do not vary 
too wildly as compared with the smallest eigenvalue. This condition seems to be quite natural as a
consideration of the diagonal case shows and furthermore allows enough
control on the smoothing of the metric to guarantee the eigenvalues,
and hence the determinant, is bounded away from zero uniformly on
compact sets for all $\eps$ small. More precisely, we have.

\begin{proposition}[Nondegeneracy of smoothed gt-regular metrics]\label{gtemb}
Let $\g$ be a nondegenerate, stable, gt-regular metric and let $\g_\eps$ be a
smoothing of $\g$ obtained by convolution with an admissible mollifier $(\rho_\eps)_\eps$.
Then its determinant $\det(\g_\eps)$ is uniformly nonvanishing on compact sets, i.e.,
\[\forall K\ \mbox{compact}\ \exists C'_K\ \exists \eps_0(K):\
 |\det(\g_\eps(x))|\geq C'_K>0\ \forall x\in K,\ \forall \eps\leq\eps_0(K).\]
In particular, the embedding $\iota_\rho(\g)$ of $\g$ is a generalised metric.
\end{proposition}

In the proof of Proposition~\ref{gtemb} we shall need the following Lemma which
exploits the stability property to give a suitable uniform control on the smoothing.
\begin{lemma}[Squeezing]\label{squeezing}
Let $f\in\Ll^\infty$ and let $L$ be compact. Suppose that there exists a
continuous function $f^L$ on $L$ such that $\|f-f^L\|_{L^\infty(L)}\leq a_L$.
Then we have
\[
\forall\ \mbox{compact}\ K\comp L\ \forall\sigma>0\  \exists\eps_0(K,\sigma):\quad\|f-f_\eps\|_{L^\infty(K)}\leq 2a_L+\sigma
\qquad\forall \eps\leq\eps_0(K,\sigma).
\]
\end{lemma}

\begin{proof}
 Let $K$, $L$ be as in the statement and write
 \begin{equation}\label{sqe}
   \|f-f_\eps\|_{L^\infty(K)}\leq\|f-f^L\|_{L^\infty(K)}+
   \|f^L-f_\eps^L\|_{L^\infty(K)}+\|f^L_\eps-f_\eps\|_{L^\infty(K)}.
 \end{equation}
 The first term on the r.h.s.\ of (\ref{sqe}) is bounded by $a_L$ and the second converges
 to zero thanks to the continuity of $f^L$. Finally, the third one is bounded by (\ref{boundedness})
 and Lemma \ref{exmol} (v) by
 \[\|f^L_\eps-f_\eps\|_{L^\infty(K)}=\|(f^L-f)*\rho_\eps\|_{L^\infty(K)}
   \leq\|f^L-f\|_{L^\infty(L)}\|\rho_\eps\|_{L^1}\leq a_L(1+\eta)
 \]
for all $\eps$ small enough.
\end{proof}

\begin{proof}[Proof of Proposition \ref{gtemb}]
 Let $K$ be compact, choose $L$ compact with $K\comp L$ and choose $\sigma$ such that
 $C''_L+\sigma/2\leq\mu_L/(2n)$. Now the stability condition (\ref{stab}) together with
 Lemma \ref{squeezing} implies that
 $\|g_{ij}-g_{ij}^\eps\|_{L^\infty(K)}\leq 2C''_L+\sigma\leq\mu_L/n$ for all $i,j$ and all
 $\eps$ small.
 Hence the maximum difference of the eigenvalues of $\g$ and $\g_\eps$ is bounded by
 \[
 \max\limits_{1\leq i\leq n}\|\lambda^i-\lambda^i_\eps\|_{L^\infty(K)}
 \leq\mathop{\mathrm{ess\,sup}}\limits_{x\in K}\|\g(x)-\g_\eps(x)\|
 \leq n\max_{1\leq i,j\leq n}\|g_{ij}-g_{ij}^\eps\|_{L^\infty(K)}\leq\mu_L\leq\mu_K,\]
 where $\|\ \|$ denotes any suitable matrix norm.
 By definition of $\mu_K$ the modulus $|\lambda^i_\eps|$ of all eigenvalues of
 $\g_\eps$ is uniformly bounded from below on $K$ for $\eps$ small enough and so is the determinant.
\end{proof}

Using the result on the determinant we are finally in a position to look at the stability
of the inverse of the smoothed metric. In particular, we have.
\begin{proposition}[Stability of the inverse]\label{inv-conv}
Let $\g$ be a nondegenerate, stable, gt-regular metric and let $\g_\eps$ be a
smoothing of $\g$ obtained by convolution with an admissible mollifier $\rho_\eps$.
Then for any open and relatively compact $V$ there exists $\eps_0(V)$ such that
the inverse of the smoothing $(\g_\eps)^{-1}$ is a smooth
and uniformly bounded net on $V$ 
for all $\eps\leq\eps_0(V)$ and we have
\[
 (\g_\eps)^{-1}\to \g^{-1}\ \mbox{in $\GTconv$ for all $p<\infty$.}
\]
In particular, for any embedding $\iota_\rho$ we have that $(\iota_\rho(\g))^{-1}\approx\g^{-1}$.
\end{proposition}

\begin{proof}
By Proposition~\ref{gtemb} $|\det(\g_\eps)|$ is locally uniformly bounded away from zero
on compact sets hence the components of the inverse of the smoothed metric $g_\eps^{ij}:=((\g_\eps)^{-1})_{ij}=\cof g^\eps_{ij}/\det(\g_\eps)$ form
a smooth and uniformly bounded net on any open, relatively compact $V$.
To prove the statement on convergence we first write
for $p<\infty$
\begin{eqnarray*}
\lefteqn{|g_\eps^{ij}-g^{ij}\|_{L^p(V)}
=\left\|\frac{\cof g^\eps_{ij}}{\det(\g_\eps)}-\frac{\cof g_{ij}}{\det\g}\right\|_{L^p}
\,=\,\left\|\frac{\cof g^\eps_{ij}\det\g-\cof g_{ij}\det(\g_\eps)}{\det\g\det(\g_\eps)}\right\|_{L^p}}\\
&\leq&\frac{1}{\|\det\g\det(\g_\eps)\|_{L^\infty}}\,
 \big(
 \|\cof g^\eps_{ij}\det\g-\cof g^\eps_{ij}\det(\g_\eps)\|_{L^p}\\
&&\hphantom{\frac{1}{\|\det\g\det(\g_\eps)\|_{L^\infty}}\,
 \big(
 \|\cof g^\eps_{ij}\det\g}
 +\|\cof g^\eps_{ij}\det(\g_\eps)-\cof g_{ij}\det(\g_\eps)\|_{L^p}
 \big)\\
 &\leq&\frac{1}{C_{\bar V}C'_{\bar V}}
 \big(
 \|\cof g^\eps_{ij}\|_{L^\infty}
 \|\det\g-\det(\g_\eps)\|_{L^p}
 +
 \|\det(\g_\eps)\|_{L^\infty}
 \|\cof g^\eps_{ij}-\cof g_{ij}\|_{L^p}
 \big),
\end{eqnarray*}
where the respective first terms are bounded by (\ref{boundedness}) and convergence 
is due to Lemma~\ref{det-conv}.

To prove $\Hl^1$-convergence we write
\begin{eqnarray*}
\lefteqn{
 \|\pa_l(g^{ij}_\eps-g^{ij})\|_{L^2(V)}}\\
 &=&
 \left\|
 \frac{\,(\pa_l\cof g^\eps_{ij})\det(\g_\eps)-\cof g^\eps_{ij}\,\pa_l\det(\g_\eps)}{(\det(\g_\eps))^2}
 -\frac{\,(\pa_l\cof g_{ij})\det\g-\cof g_{ij}\,\pa_l\det\g}{(\det\g)^2}
 \right\|_{L^2}\\
&\leq&
 \frac{1}{\|\det\g\det(\g_\eps)\|_{L^\infty}}\
 \Big(\|\det\g\,\pa_l\cof g^\eps_{ij}-\det\g\,\pa_l\cof g_{ij}\|_{L^2}\\
&&\hphantom{\frac{1}{\|\det\g\det(\g_\eps)\|_{L^\infty}}\,
 \Big(}
 {+\ \|\det\g\,\pa_l\cof g_{ij}-\det(\g_\eps)\,\pa_l\cof g_{ij}\|_{L^2}
 \Big)}\\
&&+\ \frac{1}{\|(\det\g\det(\g_\eps))^2\|_{L^\infty}}\,
 \Big(\|(\det\g)^2\cof g^\eps_{ij}\,\pa_l\det(\g_\eps)
  -(\det\g)^2\cof g^\eps_{ij}\,\pa_l\det\g\|_{L^2}\\
&&\hphantom{+\ \frac{1}{\|(\det\g\det(\g_\eps))^2\|_{L^\infty}}\,
 \Big(}
 +\ \|(\det\g)^2\cof g_{ij}^\eps\,\pa_l\det\g
  -(\det\g)^2\cof g_{ij}\pa_l\det\g\|_{L^2}\\[.3em]
&&\hphantom{+\ \frac{1}{\|(\det\g\det(\g_\eps))^2\|_{L^\infty}}\,
 \Big(}
 +\ \|(\det\g)^2\cof g_{ij}\,\pa_l\det\g
  -(\det(\g_\eps))^2\cof g_{ij}\pa_l\det\g\|_{L^2}\Big)\\
&\leq&
 \frac{1}{C'_{\bar V}}\
 \Big( \|\pa_l\cof g^\eps_{ij}-\pa_l\cof g_{ij}\|_{L^2}
      +\ \frac{1}{C_{\bar V}}\ \|\pa_l\cof g_{ij}\big(\det(\g_\eps)-\det\g\big)\|_{L^2}
 \Big)\\
&&+\ \frac{1}{{C'_{\bar V}}^2}\
 \Big( \|\cof g^\eps_{ij}\|_{L^\infty}\|\pa_l\det(\g_\eps)-\pa_l\det\g\|_{L^2}
      +\|(\pa_l\det\g)\big(\cof g^\eps_{ij}-\cof g_{ij}\big)\|_{L^2}\\
&&\hphantom{+\ \frac{1}{2{C''_{\bar V}}^2}\ \Big(}
      +\ \frac{1}{C^2_{\bar V}}\
         \|\cof g_{ij}\|_{L^\infty}\|\pa_l\det\g\big((\det\g)^2-(\det(\g_\eps))^2\big)\|_{L^2}\Big).
\end{eqnarray*}
Now the first and third term converges to zero by Lemma~\ref{det-conv} and the
bound from (\ref{boundedness}), while for the other terms we again use dominated
convergence as in (\ref{domconv}).
\end{proof}

Finally, we have a corresponding statement on the convergence of the Christoffel
symbols.
\begin{proposition}[Stability of the Christoffel symbols]\label{chr-conv}
Let $\g$ be a nondegenerate, stable, gt-regular metric and let $\g_\eps$ be a
smoothing of $\g$ obtained by convolution with an admissible mollifier $(\rho_\eps)_\eps$.
Then for any open and relatively compact $V$ there exists $\eps_0(V)$ such that
the Christoffel symbols of the first and of the second kind of the
smoothing $\Ga_{ijk}[\g_\eps]$ and $\Ga^i_{jk}[\g_\eps]$ are smooth
and $L^2$-bounded nets on $V$ for $\eps\leq\eps_0(V)$ and we have
\begin{eqnarray*}
 \Ga_{ijk}[\g_\eps]\to\Ga_{ijk}\ \mbox{and}\
 \Ga^{i}_{jk}[\g_\eps]\to\Ga^i_{jk}\ \mbox{in}\ \Ll^2
\end{eqnarray*}
In particular, for any embedding $\iota_\rho$ we have
\[
\Ga_{ijk}[\iota_\rho(\g)]\approx\Ga_{ijk}[\g]\ \mbox{and}\
\Ga^i_{jk}[\iota_\rho(\g)]\approx\Ga^{i}_{jk}[\g].
\]
\end{proposition}

\begin{proof}
Smoothness of the $\Ga_{ijk}[\g_\eps]$ is clear and $L^2(V)$-boundedness follows from
estimate (\ref{boundedness}) together with the fact that convolution commutes with taking
derivatives.

For the $\Ga^{i}_{jk}[\g_\eps]$ smoothness follows from the smoothness statement
on the inverse in Proposition~\ref{inv-conv} whereas $L^2(V)$-boundedness follows as above
and taking into account the $L^\infty(V)$-boundedness of the inverse, again given in
Proposition~\ref{inv-conv}.

As for convergence the statement on $\Ga_{ijk}[\g_\eps]$ simply follows from
Lemma~\ref{lemma:smoothing} (iii) and again the fact that the derivative of the
smoothing is the smoothing of the derivative.

For $\Ga^{i}_{jk}[\g_\eps]$ observe that we have to deal with a sum of terms of the
form $g^{ij}_\eps\,\pa_lg^\eps_{rs}=g^{ij}_\eps\,(\pa_l g_{rs})_\eps$ which due to Proposition~\ref{inv-conv} are precisely of the form covered in Lemma~\ref{basic-conv} (ii)
with $p=2$ and $m=1$.
\end{proof}

\section{Compatibility results}

We have now collected all prerequisites to precisely state our main result, saying that the
Geroch-Traschen approach to distributional metrics is compatible with the Colombeau approach.

\begin{theorem}[Compatibility for the Riemann curvature]\label{mainthm}
Let $\g$ be a nondegenerate, stable, gt-regular metric and denote its Riemann tensor
by $\Riem[\g]$. Let $\g_\eps$ be a smoothing of $\g$ obtained by convolution with an
admissible mollifier $(\rho_\eps)_\eps$. Then the we have for the Riemann tensor
$\Riem[\g_\eps]$ of $\g_\eps$
\[\Riem[\g_\eps]\to \Riem[\g]\ \mbox{in}\ \D'.\]
In other words, for any embedding $\iota_\rho(\g)$ of $\g$ we have
\[\Riem[\iota_\rho(\g)]\approx \Riem[g].\]
\end{theorem}

Before giving the proof, which using the results of the previous section is fairly
short, we illustrate the content of the theorem in a diagram.
\[
\begin{CD}
\GT\ni\g @>\iota_\rho>>[(\g_\eps)_\eps]\in\G\\
@V\mbox{$\D'$}VV @VV\mbox{Colombeau}V\\
\Riem[\g]@<\approx<<\Riem[\g_\eps]
\end{CD}
\]
Given a nondegenerate, stable and gt-regular metric $\g$ we can either derive the
Riemann curvature $\Riem[\g]$ in distributions or embed $\g$ via convolution with an admissible
mollifier to obtain the generalised metric $[(\g_\eps)]$. If we then derive its curvature
$\Riem[\g_\eps]$ within the generalised setting we find that it is associated with the
distributional curvature $\Riem[\g]$.

\begin{proof}[Proof of Theorem~\ref{mainthm}]
In coordinates we have
\[
 R^i_{jkl}[\g_\eps]=\pa_l\Ga^i_{kj}[\g_\eps]-\pa_k\Ga^i_{lj}[\g_\eps]+
 \Ga^i_{lm}[\g_\eps]\Ga^m_{kj}[\g_\eps]-\Ga^i_{km}[\g_\eps]\Ga^m_{lj}[\g_\eps].
\]

Now by Proposition~\ref{chr-conv}
$\Ga^i_{jk}[\g_\eps]\to\Ga^i_{jk}[\g]$ in $\Ll^2$ hence in $\D'$ and
we obtain $\pa_l\Ga^i_{jk}[\g_\eps]\to\pa_l\Ga^i_{jk}[\g]$ in
distributions.  By continuity of the product
$\Ll^2\times\Ll^2\to\Ll^1$ we obtain
$\Ga^i_{jk}[\g_\eps]\Ga^l_{rs}[\g_\eps]\to\Ga^i_{jk}[\g]\Ga^l_{rs}[\g]$
in $\Ll^1$, hence again in distributions.

\end{proof}

Similarly we also have compatibility results for the Ricci-, Weyl- and
scalar curvature. To prepare for the formulation and proof of these
results we recall from \cite{gt} that it is possible to
define the outer product of (any number of copies of inverses of) a
gt-regular metric with its Riemann tensor.  Indeed in the smooth case
we may write
\[
\frac{1}{2}\,g^{rs}R^i_{jkl}=g^{rs}(\pa_{[l}\Ga^i_{k]j}+g^{rs}\Ga^i_{m[l}\Ga^m_{k]j})
=\pa_{[l}\big(g^{rs}\Ga^i_{k]j}\big)-(\pa_{[l}g^{rs})\Ga^i_{k]j}+g^{rs}\Ga^i_{m[l}\Ga^m_{k]j},
\]
and we see that the right hand side makes sense in distributions for a
gt-regular metric. Indeed, $g^{rs}\Ga^i_{kj}\in\Ll^2$ allows for a
weak derivative as well as $(\pa_{l}g^{rs})\Ga^i_{kj}\in\Ll^1 \ni
g^{rs}\Ga^i_{ml}\Ga^m_{kj}$. Moreover, the same holds true for any product of the
form $\otimes_m\g\otimes_l\g^{-1}\otimes\Riem[\g]$: just use the
Leibnitz rule on $\pa_l(\otimes_m\g\otimes_l\g^{-1}\Ga^i_{jk})$. We now have.

\begin{corollary}[Compatibility for curvature quantities]
Let $\g$ be a nondegenerate, stable, gt-regular metric and
let $\g_\eps$ be a smoothing of $\g$ obtained by convolution with an
admissible mollifier $(\rho_\eps)_\eps$. Then the we have ($m,l\in\N$)
\[
\otimes_m\g_\eps\otimes_l\g_\eps^{-1}\otimes\Riem[\g_\eps]\to\otimes_m\g\otimes_l\g^{-1}
\otimes\Riem[g]\ \mbox{in}\ \D'.
\]
In particular, the result applies to the Ricci-, Weyl- and scalar curvature and with other words
we have for any embedding $\iota_\rho(\g)$ of $\g$
\[
\Ric[\iota_\rho(\g)]\approx\Ric[\g],\ W[\iota_\rho(\g)] \approx W[\g],\ R[\iota_\rho(\g)]\approx R[\g].
\]
\end{corollary}

\begin{proof}
According to the above discussion we have to deal with the terms
$$\pa_l\Big(g^\eps_{i_1j_1}\cdots g^\eps_{i_mj_m}g^{r_1s_1}_\eps\cdots g^{r_ls_l}_\eps\Ga^i_{jk}[\g_\eps]\Big),\
\pa_l\Big(g^\eps_{i_1j_1}\cdots g^\eps_{i_mj_m}g^{r_1s_1}_\eps\cdots g^{r_ls_l}_\eps\Big)\Ga^i_{jk}[\g_\eps]$$
and
$$g^\eps_{i_1j_1}\cdots g^\eps_{i_mj_m}g^{r_1s_1}_\eps\cdots g^{r_ls_l}_\eps\Ga^i_{mj}[\g_\eps]\Ga^m_{lj}[\g_\eps].$$
To deal with the first one note that
\[
g^\eps_{i_1j_1}\cdots g^\eps_{i_mj_m}g^{r_1s_1}_\eps\cdots g^{r_ls_l}_\eps\Ga^i_{jk}[\g_\eps]\to g_{i_1j_1}\cdots g_{i_mj_m}g^{r_1s_1}\cdots g^{r_ls_l}\Ga^i_{jk}[\g]
\]
in $\Ll^2$ by Lemma~\ref{basic-conv}(ii) for $p=2$, hence in distributions and we obtain
the desired convergence
of the derivatives. For the second term note that by the Leibnitz rule we only have to show that
\[
g^\eps_{i_1j_1}\cdots(\pa_lg^\eps_{i_pj_p})\cdots g^\eps_{i_mj_m}g^{r_1s_1}_\eps\cdots g^{r_ls_l}_\eps
\Ga^i_{jk}[\g_\eps]
\to
g_{i_1j_1}\cdots(\pa_lg_{i_pj_p})\cdots g_{i_mj_m}g^{r_1s_1}\cdots g^{r_ls_l}\Ga^i_{jk}[\g]
\]
(and analogously for the terms with the derivative falling on the inverse). However, this holds
true in $\Ll^1$, hence $\D'$ by Lemma~\ref{basic-conv}(ii) and by continuity of the product 
$\Ll^2\times\Ll^2\to\Ll^1$. Finally, the same argument applies
to $g^\eps_{i_1j_1}\cdots g^\eps_{i_mj_m}g^{r_1s_1}_\eps\cdots g^{r_ls_l}_\eps\Ga^i_{mj}[\g_\eps]\Ga^m_{lj}[\g_\eps]$.
\end{proof}

Finally, we discuss the relation of our results to the stability
results obtained by Geroch and Traschen in \cite{gt} and LeFloch and
Mardare in \cite{LFM}. To begin with we remark that in their Theorem
4.6, LeFloch and Mardare \cite{LFM} suppose convergence of
$\g^{-1}_\eps$ to $\g^{-1}$ in $\Ll^\infty$ which is not true in case
of smoothings via convolution unless the metric is supposed to be more
regular, e.g.\ continuous. In this case our result coincides with
theirs while in general we deal with nets that converge only
in a weaker sense.

On the other hand the relation with the results of Geroch and Traschen
is more subtle.  Theorem 2 of ~\cite{gt} asserts that for any sequence
of gt-regular metrics $\g_n$ that is $\Ll^\infty$-bounded together
with its inverse $(\g_n)^{-1}$ and for which $\g_n$, $(\g_n)^{-1}$ and
$\pa_i\g_{n}$ converge in $\Ll^2$ to $\g$, $\g^{-1}$ resp.\ $\pa_i\g$
the sequence $\Riem[\g_n]$ of Riemann tensors converges to $\Riem[\g]$
in $\D'$.  Actually, in the context of the present work, the
nondegenerate and stability conditions we impose on a gt-regular metric
ensures that the conditions required for their Theorem 2 are satisfied
as a consequence of our Lemma~\ref{lemma:smoothing} together with
Propositions ~\ref{inv-conv} and \ref{chr-conv}. Thus our Theorem
\ref{mainthm} follows from \cite[Thm.\ 2]{gt} but we feel that our
proof is more direct.  Indeed a mild variation of our proof provides
a simpler proof of their Theorem.  

Also note that our results on the stability of the inverse metric and
the Christoffel symbols are more precise and actually provide the
best possible $\Wl^{m,p}$-convergence: If we had converge in any
smaller $\Wl^{m,p}$-space then by completeness the original metric
would have had to be in that space too.

Finally, we note that Theorem 4 in \cite{gt} shows that for
any continuous, gt-regular metric there is a sequence of smooth
metrics (actually obtained by smoothing via convolution) which
converges in the above mentioned sense. However, recall from the discussion preceding
Definition \ref{gt-stab} that the question of
nondegeneracy in the continuous case is much easier to handle. The
question of whether the requirement for continuity could be omitted from
the assumptions was left open in \cite{gt} with the proof failing to
cover this case. Our results provide a positive answer to this
question: For any nondegenerate, stable, gt-regular metric the
smoothing provides a smooth sequence which converges in the desired
sense.

\begin{appendix}
\section{The existence of suitable mollifiers}
As pointed out in Section~\ref{colombo} above one crucial feature of
the Colombeau approach is that the space $\Cinfty$ of smooth functions
is a subalgebra of the algebra of generalised functions $\G$. This is
achieved by the fact that the embeddings $\iota$ and $\sigma$ coincide
for smooth functions, i.e., $\sigma(f)-\iota(f)\in\ns$ for all smooth
$f$. The crucial estimate (cf.\ \cite[Prop.\ 1.2.11]{GKOS}) in turn is
based on the fact that the mollifier used to define $\iota$ is assumed
to have vanishing moments. It is actually this requirement that forces
us to assume $\rho\in{\mathcal S}$ since there exist no compactly
supported smooth functions with all moments vanishing. Moreover, a
function with all moments vanishing can never be nonnegative and also
has an infinite number of zeroes.  It is this property which makes it
a nontrivial task to preserve positivity when embedding distributions
into $\G$. One solution to this problem is discussed in this appendix.

The key step in our approach is to replace the embedding $\iota$ by
convolution with a suitable strict $\delta$-net $(\rho_\eps)_\eps$
which eventually has vanishing moments and has negative part with
arbitrary small $L^1$-norm. Since we are now convolving with a strict
$\delta$-net rather than a model $\delta$-net, i.e., a net obtained by
scaling a single function $\vphi$, we have to be careful to obtain
moderateness of $(u*\psi_\eps)_\eps$ (cf.\
\cite[Prop. 1.2.10]{GKOS}). The latter property will be a consequence
of moderateness of $(\rho_\eps)_\eps$ itself. We start by providing a
suitable net $(\psi_\eps)_\eps$: the scaled version denoted
by $(\rho_\eps)_\eps$ being the  admissible mollifiers used in
section~\ref{smoothing gt}.

\begin{lemma}[Existence of suitable mollifiers]\label{exmol1}
 There exists a net $(\psi_\eps)_\eps$ of test functions on $\R^n$ with the properties
 \begin{itemize}
  \item [(i)] $\displaystyle\supp(\psi_\eps)\subseteq B_1(0)$\quad for all $\eps\in(0,1]$
  \item [(ii)] $\displaystyle\int\psi_\eps(x)\, dx=1$\quad for all $\eps\in(0,1]$
  \item [(iii)] $\displaystyle\forall\alpha\in\N_0^n\ \exists p:\
              \sup_{x\in\R^n}|\partial^\alpha\psi_\eps(x)|=O(\eps^{-p})$
  \item [(iv)] $\displaystyle\forall j\in\N\ \exists\eps _0:\
              \int x^\alpha\psi_\eps(x)\,dx=0$\quad for all $1\leq|\alpha|\leq j$ and all $\eps\leq\eps_0$
  \item [(v)] $\displaystyle\forall \eta>0\ \exists \eps_0:\
              \int|\psi_\eps(x)|\,dx\leq 1+\eta$\quad for all $\eps\leq\eps_0$.
 \end{itemize}
 In particular,
 \[\rho_\eps:=\frac{1}{\eps^n}\,\psi_\eps\Big(\frac{.}{\eps}\Big)\]
 is a strict $\delta$-net, which is moderate, has finally vanishing moments and its negative
 parts have arbitrarily small $L^1$-norm, i.e., $\rho_\eps$ satisfies (iii)--(v).
\end{lemma}

This statement can actually be proved by an application of \cite[Thm.\ 3.10]{OV} along the
lines of \cite[Props. 5.1, 5.2]{OV}. However, since this reference uses the language of
``internal sets''---a concept inspired by nonstandard analysis (for related work see also
\cite{OT})---we have chosen to include a direct proof.

\begin{proof}
 We will be concerned with the following sets ($m\in\N_0$, $\eta>0$)
 \begin{eqnarray*}
  \A_m&\!:=\!&\{\vphi\in\D(\R^n):\, \supp(\vphi)\subseteq B_1(0),\, \int\vphi=1,\
            \int x^\al\vphi(x)\,dx=0\quad\forall 1\leq|\al|\leq m\},\\
  \A'_m(\eta)&\!:=\!&\{\vphi\in\A_m:\ \int|\vphi|\leq1+\eta\}.
 \end{eqnarray*}
 It is well known that the sets $\A_m\not=\emptyset$ (see e.g.\ \cite[Pro.\ 1.4.2]{GKOS}; the
 additional requirement on the supports asserted here is easily obtained by scaling).
 Following \cite[Prop.\ 5.1.]{OV} we now prove that also the sets
 \[ \A'_m(\eta)\not=\emptyset,\quad\mbox{for all $m\in\N_0$ and all $\eta>0$}.\]
 It suffices to prove the result in the 1-dimensional case $n=1$: the
 general case then follows by taking tensor products of functions of one
 variable. We proceed by induction.\\ \underline{$m=0$}:
 $\A'_m(\eta)\not=\emptyset$ even for $\eta=0$, since it suffices to choose
 $0\leq\vphi \in\D(\R)$ with $\supp(\vphi)\subseteq B_1(0)$ and
 $\int\vphi=1$.\\ \underline{$m-1\mapsto m$}: Let
 $\vphi\in\A'_{m-1}(\eta/2)$ and set $\psi:=a\vphi+b\vphi(./\mu)$,
 where $a$, $b$, and $0<\mu<1$ are to be specified below. We have
 \[\int\psi=a+b\mu,\ \int x^k\psi(x)\,dx=0\quad\forall 1 \leq k\leq m-1,\]
 as well as
 \[\int x^m\psi(x)\,dx=(a+b\mu^{m+1})\int x^m\vphi(x)\,dx.\]
 Solving $a+b\mu=1$ and $a+b\mu^{m+1}=0$ for $a$ and $b$ we obtain
 \[ a=\frac{-\mu^m}{1-\mu^m}\ (<0)\ \mbox{and}\ b=\frac{1}{\mu-\mu^{m+1}}\ (>0)\]
 and so
 \[\int|\psi|\leq(|a|+|b|\mu)\int|\vphi|\leq\frac{1+\mu^m}{1-\mu^m}\ \Big(1+\frac{\eta}{2}\Big),\]
 which can be made smaller than $1+\eta$ if $\mu$ is chosen small enough. So we obtain
 $\psi\in\A'_m(\eta)$ and we are done.
 \medskip

 Now we choose
 \[ \vphi_m\in\A'_m(1/m)\ \mbox{and set}\ M_m:=\sup_{x\in\R^n,|\al|\leq m}|\pa^\al\vphi_m(x)|\]
 and define the sets 
 \[\A_{m,\eps}:=\{\vphi\in\A'_m(1/m):\ \sup_{x\in\R^n,|\al|\leq m}|\pa^\al\vphi(x)|\leq\frac{1}{\eps}\}.\]
 Note that by the above $\A_{m,\eps}\not=\emptyset$ if $\eps\leq 1/M_m=:\eps_0(m)$ and
 $\A_{m+1,\eps}\subseteq\A_{m,\eps}$ for all $\eps$. Now for
 $m\in\N$, $\eps\leq\eps_0(m)$ we choose $\psi_{m,\eps}\in\A_{m,\eps}$ and finally set
 \[\psi_\eps:=\psi_{m,\eps}\qquad \eps_0(m+1)<\eps\leq\eps_0(m).\]

 We then obviously have (i) and (ii) and it remains to verify (iii)-(v).\\
 (iii): Let $|\al|\in\N^n_0$. Then since $\psi_\eps\in\A_{|\al|,\eps}$ for $\eps\leq\eps_0(|\al|)$ 
   we obtain
   $\sup_{x\in\R^n}|\pa^\al\psi_\eps(x)|\leq1/\eps$ for all such $\eps$.\\
 (iv): Let $|\al|\geq 1$. Then since $\psi_\eps\in\A_{|\al|}$ for $\eps\leq\eps_0(|\al|)$ we have
   $\int x^\al\psi_\eps(x)dx=0$ for all such $\eps$.\\
 (v): Let $\eta>0$ and choose $m$ such that $1/m\leq\eta$. Since $\psi_\eps\in\A'_m(1/m)$ for all
   $\eps\leq\eps_0(m)$ we have for all such $\eps$ that $\int|\psi_\eps|\leq 1+1/m\leq 1+\eta$.\\
  \end{proof}

Finally we observe that the mollifiers obtained above in fact provide an embedding of distributions
into the Colombeau algebra.

\begin{corollary}[An embedding of distributions]\label{ca1}
Let $u\in\D'(\R^n)$ and let $(\rho_\eps)_\eps$ be a strict $\de$-net as in Lemma \ref{exmol1}.
Then the mapping
\[\iota_\rho:\ u\mapsto [(u*\rho_\eps)_\eps]\]
is a linear embedding of $\D'(\R^n)$ into $\G(\R^n)$ having the distinguishing
properties
\begin{enumerate}
\item $\iota_\rho\circ\pa^\al=\pa^\al\circ\iota_\rho$\quad for all $\al\in\N_0^n$
\item $\iota_\rho|_{\Cinfty}=\sigma$
\item $\iota_\rho(u)\approx u$
\item $\iota_\rho$ preserves supports.
\end{enumerate}
\end{corollary}

\begin{proof} The proof is just a mild variation of the usual ``standard proofs''.
So we only remark that for proving moderateness of $\iota_\rho(u)$ as well as for proving
(ii) and (iv) ((i) and (iii) follow directly from the properties of the convolution)
we just have to use moderateness of $(\rho_\eps)_\eps$ in the respective proofs
of Propositions 1.2.10--1.2.12 in \cite{GKOS}.
\end{proof}

\end{appendix}


\end{document}